\documentclass[sigconf]{acmart}

\AtBeginDocument{%
  \providecommand\BibTeX{{%
    \normalfont B\kern-0.5em{\scshape i\kern-0.25em b}\kern-0.8em\TeX}}}

\copyrightyear{2024}
\acmYear{2024}
\setcopyright{rightsretained}
\acmConference[PEARC '24]{Practice and Experience in Advanced Research Computing}{July 21--25, 2024}{Providence, RI, USA}
\acmBooktitle{Practice and Experience in Advanced Research Computing (PEARC '24), July 21--25, 2024, Providence, RI, USA}
\acmDOI{10.1145/3626203.3670536}
\acmISBN{979-8-4007-0419-2/24/07}

\acmConference[PEARC '24]{Practice and Experience in Advanced Research Computing}{July 21--25,
  2025}{Providence, RI}

\begin{document}

\title{Benchmarking with Supernovae: A Performance Study of the FLASH Code}

\author{Joshua Martin}
\authornote{Both authors contributed equally to this research.}
\email{joshua.martin.1@stonybrook.edu}
\orcid{0009-0000-4674-0192}
\author{Catherine Feldman}
\email{catherine.feldman@stonybrook.edu}
\orcid{0000-0001-7003-4431}
\authornotemark[1]
\author{Eva Siegmann} 
\email{eva.siegmann@stonybrook.edu}
\orcid{0000-0003-1216-1576}
\author{Tony Curtis}
\email{anthony.curtis@stonybrook.edu}
\orcid{0000-0002-8192-0700}
\author{David Carlson}
\email{david.carlson@stonybrook.edu}
\orcid{0000-0002-6987-9493}
\author{Firat Co\c skun}
\email{firat.coskun@stonybrook.edu}
\orcid{0009-0006-6755-8772} 
\author{Daniel Wood}
\email{daniel.wood@stonybrook.edu}
\orcid{0009-0003-7271-3923}
\author{Raul Gonzalez}
\email{raul.gonzalez@stonybrook.edu}
\orcid{0009-0006-3071-2473} 
\author{Robert J.\ Harrison}
\email{robert.harrison@stonybrook.edu}
\orcid{0000-0002-8777-7466}
\author{Alan C.\ Calder}
\email{alan.calder@stonybrook.edu}
\orcid{0000-0001-5525-089X}

\affiliation{%
  \institution{Institute for Advanced Computational Science}
  \streetaddress{Stony Brook University}
  \city{Stony Brook}
  \state{New York}
  \country{USA}
  \postcode{11794-5250}
}

\renewcommand{\shortauthors}{Martin and Feldman, et al.}

\begin{abstract}
Astrophysical simulations are computation, memory, and thus energy intensive, thereby requiring new hardware advances for progress. Stony Brook University recently expanded its computing cluster “SeaWulf” with an addition of 94 new nodes featuring Intel Sapphire Rapids Xeon Max series CPUs. We present a performance and power efficiency study of this hardware performed with FLASH: a multi-scale, multi-physics, adaptive mesh-based software instrument. We extend this study to compare performance to that of Stony Brook’s Ookami testbed which features ARM-based {\color{black} A64FX-700} processors, and SeaWulf’s AMD EPYC Milan and Intel Skylake nodes. Our application is a stellar explosion known as a thermonuclear (Type Ia) supernova and for this 3D problem, FLASH includes operators for hydrodynamics, gravity, and nuclear burning, in addition to routines for the material equation of state. We perform a strong-scaling study with a 220 GB problem size to explore both single- and multi-node performance.
Our study explores the performance of different MPI mappings and the distribution of processors across nodes. From these tests, we determined the optimal configuration to balance runtime and energy consumption for our application.
\end{abstract}





\maketitle

\section{Introduction}

Computational science advances in a host of ways, including improved algorithms, techniques for handling burgeoning data, and improvements in computational hardware. Newer CPU architectures take advantage of non-uniform memory access (NUMA) patterns, advances in fabrication that allow for a higher per-node core count and larger caches, and new types of memory such as high-bandwidth memory (HBM). A significant advantage for application scientists is the ability to use these new features without substantial rewriting of code. Two new architectures designed for this purpose are Intel's Xeon Max with HBM and Fujitsu's A64FX. This study explores the performance on these architectures of FLASH, an established community code with a significant and diverse user base \cite{Dubey_2019}.

 FLASH is a multi-scale, multi-physics simulation instrument originally designed to model astrophysical reactive flows
and extended to study high-energy-density environments~\cite{Fryxell_2000,flashhed}. 
FLASH was created at the 
University of Chicago's Flash Center, as part of the DOE's Accelerated Strategic Computing Initiative (ASCI)~\cite{asci2000}. 
The ASCI program and
its follow-on, the ASC program, allowed
development for advanced architectures as they came online at the
national laboratories, thereby establishing a long
history of use on state-of-the-art platforms. 
Accordingly, development has not been optimized for any one architecture.
A highlight from FLASH's evolution was
achieving 238 GFlops on 6420 processors of ASCI Red at the
Los Alamos National Laboratory while performing adaptive mesh refinement
simulations of reactive flow, which won the SC2000 Gordon Bell 
prize~\cite{calder.curtis.ea:high-performance}. 

Recently FLASH
performed the largest-ever simulation of supersonic turbulence
on the SuperMUC system. The simulation had
an effective resolution of 10,048$^3$ cells and ran
on 65,536 Intel Xeon E5-2680 (Sandy Bridge) cores for 50 million CPU hours~\cite{Turbulence_Federrath_2021}.
FLASH was also one 
of the initial applications for the testbed cluster
Ookami~\cite{ookamiurl,ookamixsede} that provides open access to 
a testbed supercomputer featuring {\color{black}A64FX-700} Fujitsu compute 
nodes~\cite{flash_huge_2022,flash_huge_2023}. 
Development of FLASH for astrophysics continues~\cite{townsleyetal2019},
and a new code FLASH-X, derived from FLASH with a completely new infrastructure, is being developed
for performance portability across heterogeneous (CPU + GPU) architectures~\cite{Oneal2018}.
The present FLASH
we used for this study is parallelized primarily through MPI, however, and is thus largely bound to CPU architectures. 
These demonstrated capabilities make FLASH a perfect candidate to investigate the 
merit of new CPU architectures for real-world applications.

Our application is a 3D nuclear fusion driven explosion known as a
thermonuclear (Type Ia) supernova.  A contemporary introduction to
these events and further details of our 2D simulation can be found in~\citet{feldmanetal2023a}.
We use FLASH version 4.6.2 
including the PARAMESH library to manage a block-structured adaptive 
mesh~\cite{macneice.olson.ea:paramesh, macneice.olson.ea:paramesh*1}.
PARAMESH Morton-orders the blocks for
communication efficiency and load balancing, and the maximum numbers
of blocks per processor, hence the maximum memory used per processor, is
set by the parameter \texttt{maxblocks}.
Each 3D block consists of 16$^3$ cells and guard cells, which contain copies of data from 4 cells deep of the neighboring blocks. 
Each cell is described by 29 cell-centered variables such as density,
temperature, and pressure. The data container for these variables is a single array for the entire block, ordered by cell rather than by variable type. This means that accessing the same variable over different cells introduces a memory stride of 29 floating point numbers. 

At each timestep, FLASH first calls the selected physics solvers, which for our problem are a compute-heavy, local communication hydrodynamics solver;
a global communication, multipole expansion approximate Poisson solver
for gravity; and other related solvers.
FLASH then determines if the mesh needs to be refined
and, if so, adjusts the number of blocks, interpolates or copies the cell data, and redistributes the blocks amongst the processors. Each solver may perform differently on different systems.


\section{Hardware and Software Setup}

\subsection{Hardware}
We have access to two compute clusters: "SeaWulf", a CPU-based cluster with nodes of different architectures that has recently expanded to include Intel's Sapphire Rapids with HBM nodes (SPR), and "Ookami", an A64FX testbed. The goal of this expansion and testbed is to enable increased performance for codes with a traditional programming model for CPUs.
The features of nodes explored in this study is shown in Table~\ref{tab:CPUarch}. Hyperthreading, where available, is disabled.


\begin{table*}[ht!]
\caption{Features of the different types of nodes used in this study. Information was gathered from \texttt{lscpu} and \texttt{cpupower frequency-info}.}
\label{tab:CPUarch}
\begin{tabular}{llllllllll}
\toprule
CPU Architecture & \begin{tabular}[c]{@{}l@{}}Launch \\ Date \end{tabular} & Sockets & \begin{tabular}[c]{@{}l@{}}NUMA \\ Regions\end{tabular} & Cores & Freq. (GHz) & RAM (GB) & LLC (MB) & InfiniBand \\
\midrule
\begin{tabular}[c]{@{}l@{}}Intel Xeon Max 9468 \\ "Sapphire Rapids w/ HBM"\end{tabular} & {\color{black}2023} & 2 & 8 & 96 & 0.80 - 3.5 & \begin{tabular}[c]{@{}l@{}}256 DDR5\\ 128 HBM\end{tabular} & \begin{tabular}[c]{@{}l@{}}L3, 105 \\ (48 cores)\end{tabular} & NDR (400 Gb/s) \\
\hline
\begin{tabular}[c]{@{}l@{}}Intel Xeon Gold 6148\\ "Skylake"\end{tabular} & {\color{black}2017} & 2 & N/A & 40 & 2.4 & 192 DDR4 & L3, 27.5  & FDR (56 Gb/s) \\
\hline
\begin{tabular}[c]{@{}l@{}}AMD EPYC 7643 \\ "Milan"\end{tabular} & {\color{black}2021} & 2 & 8 & 96 & 1.5, 1.9, 2.3 & 256 DDR4 & \begin{tabular}[c]{@{}l@{}}L3, 32\\ (6 cores)\end{tabular} & HDR (100 Gb/s) \\
\hline
Fujitsu A64FX-700 & {\color{black}2019} & 1 & 4 & 48 & 1.8 & 32 HBM & \begin{tabular}[c]{@{}l@{}}L2, 8\\ (48 cores)\end{tabular} & HDR (100 Gb/s) \\
\hline
\end{tabular}
\end{table*}

For the SPR nodes, SeaWulf uses the HBM Flat Mode configuration with sub-NUMA clustering SNC4. This means that the  HBM is used as additional RAM to the DDR5. The SPR nodes' BIOS setting have been customized to optimize energy effiency: the CPU frequency is continuously updated by the Dynamic Power Savings Power Regulator with a Balanced Performance Energy/Performance Bias and Energy Efficient Turbo enabled, and the Minimum Processor Idle Power Core C-state is set to C6.


\subsection{Software}
The software, compiler flags, and MPI options used for each architecture are shown below in Table~\ref{tab:flash_compilers}. Unfortunately, Ookami no longer has a license for the Fujitsu compiler and therefore it could not be used; however, the Fujitsu compiler has been shown to provide superior performance for FLASH~\cite{flash_huge_2023}. FLASH requires double-precision reals and doubles, and single-precision integers --- additional compiler flags selecting those options were included. We use the HDF5 1.12.1 library for parallel I/O. Using the HBM on the SPR nodes requires loading the numactl 2.0.16 module and adding \texttt{numactl --preferred-many=8-15} to the \texttt{mpiexec} command at runtime.


\begin{table*}[ht!]
  \caption{Compiler, MPI versions, and compiler flags used for each architecture}
  \centering
  \begin{tabular}{l l c c l}
    \hline \hline
    Architecture & Compiler & MPI & Additional Compiler Flags \\
    \hline
     SPR & GCC 12.1 & OpenMPI 4.1.5 & \texttt{ -Ofast}\\
     & & & \texttt{-march=sapphirerapids -mtune=sapphirerapids}  \\
     \hline
     SPR & Intel 2023.1 & Intel MPI 2021.9.0 & \texttt{ -Ofast -xsapphirerapids}\\
     \hline
     Milan & GCC 12.1 & OpenMPI 4.1.4 & \texttt{ -Ofast -march=native}\\
     \hline
     Milan & AOCC 4.0 & OpenMPI 4.1.5 & \texttt{ -Ofast -march=native}\\
     \hline
     Skylake & GCC 12.1 & OpenMPI 4.1.4 & \texttt{ -Ofast -march=skylake -mtune=skylake}\\
     \hline
     Skylake & Intel 2023.1 & Intel MPI 2021.9.0 & \texttt{ -Ofast {\color{black} -march=skylake -mtune=skylake} }\\
     \hline
     A64FX & GCC 12.1 & OpenMPI 4.1.4 & \texttt{ -Ofast {\color{black} -mcpu=a64fx}}\\
     \hline
  \end{tabular}
  \label{tab:flash_compilers}
\end{table*}



{\color{black} We selected a problem size for our supernova application that fit within the DDR5 of an SPR node -- $\sim$220 GB when using 96 nodes on SPR+GCC -- to be able to test single-node as well as multi-node performance.
We explored strong scaling of our application on all systems (Section~\ref{sec:1node}),
and additional tests were then performed based on their results including 
better understanding the SPR HBM (Section~\ref{sec:hbm}) and
spreading the problem
across nodes (Section~\ref{sec:corespread}),} 

\subsection{Data Collection}
\subsubsection{Runtime}
Unless otherwise specified, we run each simulation 7 times on the same (group of) nodes, and report the average and standard deviation of the evolution runtime. 
The evolution runtime excludes the initialization and finalization of reading and writing large checkpoint files, so the architectures can be better compared on the application performance rather than I/O. 
FLASH also collects runtime data separately for each solver (e.g. hydro, gravity, nuclear burning, grid refinement).
FLASH provides information about the maximum, minimum, and average runtime over all of the processors running the application. Some processors take longer than others to finish due to MPI collectives, communication, or the complexity of the calculations for a certain block (ie the Riemann solve might take more iterations to converge for one block than for another).
Unless otherwise specified, we report the maximum runtime along with its standard deviation.

\subsubsection{Power and Energy}
On SeaWulf, power is sampled once per minute by \texttt{ipmitool}. To obtain power and energy use measurements, we calculated the average power for each of the 7 runs, and multiplied by the corresponding runtime. 
When calculating the average power, the first and last data points are excluded to remove any warm-up/cool-down that the processor undergoes at the beginning/end of each run. Including these skews the average power values to be slightly lower. The average and standard deviation of these 7 energy measurements are reported.
On Ookami, power and energy data is accessed through {\color{black} Open XDMoD \citep{xdmod}. The average single-node power is reported in XDMoD for each job (in this case, the average over 7 runs). As the CPU frequency is constant on A64FX, the power draw is very stable, and doesn't change much between runs. To calculate the energy for a single run, we multiply the average power for each job by the number of nodes used and the runtime. As with SeaWulf, the average and standard deviation of these 7 energy measurements are reported.}

{\color{black}\section{Scaling Study}} \label{sec:1node}
This 3D problem contains 33,016 16 $\times$ 16 $\times$ 16 blocks, and is run for 20 timesteps including 10 grid refinements. At the end of the run, the number of blocks has only modestly increased to 33,864, allowing the memory use to remain nearly constant over the course of the simulation. To keep the memory use consistent as the number of processors changes, we allocate a total of 37,000 blocks, adjusting \texttt{maxblocks} accordingly for all architectures but {\color{black}A64FX-700}, which required a \texttt{maxblocks} of 50. The number of timesteps was selected so that a 96-core SPR run would take roughly 20 minutes, which should provide enough data for a power measurement.

\subsection{SeaWulf Cluster}
Before revealing the strong-scaling abilities of all configurations, it is worth noting a few caveats in our results. First, we were unable to run our application on just 1 or 2 cores, so our strong-scaling study starts at 4 cores. There are missing runs in both the Intel-SPR runs as well as the Skylake runs due to miscellaneous errors such as segmentation faults and issues computing our equation of state. It should also be noted that Skylake nodes only contain 40-cores unlike Milan and SPR's 96-core nodes. This meant that our largest Skylake run of 320 cores spanned 8 Skylake nodes.

\begin{figure}[ht!]
\centering
   \includegraphics[width=0.5\textwidth]{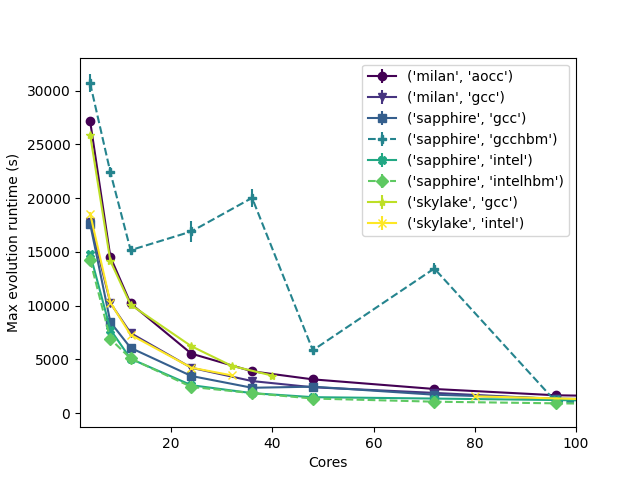}
   \includegraphics[width=0.5\textwidth]{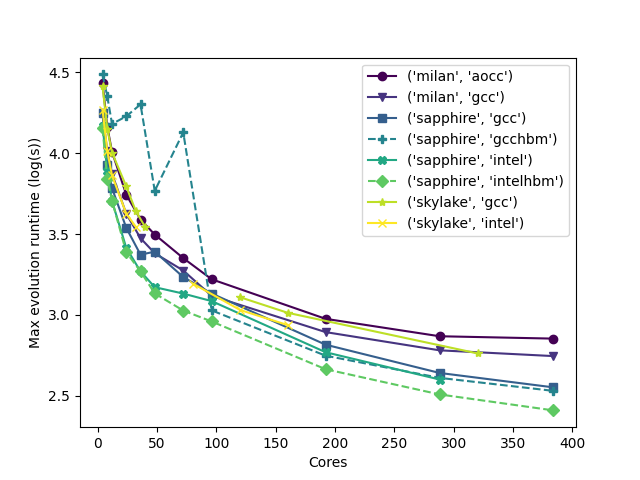}
  \vspace{-15pt} 
  \caption{\label{fig:small_strong_scaling} Top panel: Strong-scaling studies on all configurations available on SeaWulf from 4 to 96 cores (single node). Missing data points are explained above.
Bottom panel: The same strong-scaling study but expanded out to 384 cores and expressed as a log plot.}
\end{figure}

Available in Figure~\ref{fig:small_strong_scaling} is the full strong-scaling study spanning from 4 to 384 cores. We chose to extend this study to multiple nodes to assess cross-node communication efficiency. An immediate takeaway from this study are that switching from GCC to the native compiler for a given node did not always result in a speedup. While Intel did show this speedup when switching to its native compiler, AMD did not as evidenced by its native AOCC falling well short of the times posted by running Milan with GCC. It is also clear that scaling begins to considerably slow past the 192-core mark for all configurations. Past this 192-core mark, the difference in scaling ability between SPR nodes and our older architectures becomes most apparent as the older architectures demonstrate a larger stagnation in decline.



\begin{figure}
    \centering
    \includegraphics[width=0.47\textwidth]{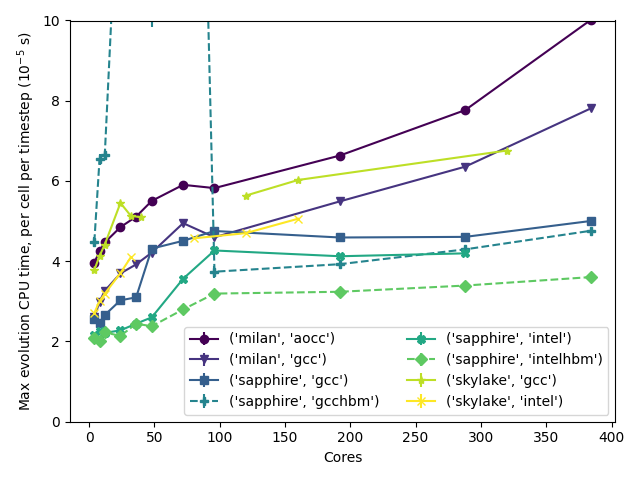}
    \vspace{-15pt}
    \caption{Per-cell per-timestep CPU time on all configurations available on SeaWulf from 4 to 384 cores. Missing data points are explained above.}
    \label{fig:small_dean}
\end{figure}

Figure~\ref{fig:small_dean} further showcases the difference in scaling ability between SPR and older architectures. In Figure~\ref{fig:small_dean}, we look at how long it takes for a single section of our simulation, called a cell, to be fully computed. Specifically, we view the maximum cell CPU time per timestep (MCCTT) as the simulation will not advance in timestep until all cells are computed. Therefore, our slowest cell computation on each timestep is most relevant to scaling ability. The MCCTT is calculated by dividing the maximum evolution CPU time by the number of timesteps and the average number of cells (average number of blocks $\times$ 16$^3$ cells per block).

In perfect strong-scaling, the number of CPU hours, and therefore the MCCTT, remains constant over core number. The speedup in our strong-scaling purely derives from more of the cells being simultaneously computed due to increased total core count. It then follows that the MCCTT will only increase as the number of cores are increased due to increased communication time. This is because the core responsible for a given cell's computation will need to receive information from a greater number of cores (which will begin to be further away in physical space) responsible for nearby (in simulation space) cells. In expanding our computation to multi-node, these communications also begin to face the issue of limited inter-node communication bandwidth.

In the multi-node portion of Figure \ref{fig:small_dean}, it is clear that SPR faces a much more benign slowdown in MCCTT than the older configurations. This is can be immediately traced back to the node-interconnect "InfiniBand" bandwidths being much larger on SPR. Whereas SeaWulf's Milan nodes possess a node-interconnect bandwidth of 100 GB/s, the corresponding bandwidth on SPR is 400 GB/s.

However, in the single-node portion of Figure~\ref{fig:small_dean}, we notice a peculiar increase in the MCCTT of all configurations. We discovered that, as we begin to utilize more cores on a given node, the CPU frequency of that core decreases. This explains why the SPR plots are only flat past the single-node mark. This issue is discussed more at length in Section~\ref{sec:corespread}.

Another point of discussion is the HBM results. Barring Ookami (discussed in Section~\ref{sec:ookami}), it was unsurprising that SPR, our newest, most advanced architecture, using its native compiler with HBM enabled recorded the fastest runtimes. Following close behind to these runtimes are the higher core runs of SPR with GCC and HBM enabled. However, the sub-single-node runs of this configuration posted unusually slow runtimes. We further explore this anomaly in Section~\ref{sec:hbm}{\color{black}, which appears to be related to MPI process mapping}.

With all configurations seeing a significant decrease in scaling ability past the 192-core mark, it is natural to ask what the bottleneck of our simulation is. Activating the HBM does not provide the substantial speedup expected for memory-bound applications. {\color{black} Additionally, \citet{EvaSPR} demonstrated that an SPR node's DDR5 memory will become saturated when using only 24 cores. Good scaling of our application continues past 24 cores per node, meaning the saturation of DDR5 does not negatively impact runtime and our application is not memory-bound.} Investigating the runtime breakdown of our application shows that FLASH spends over 50\% of the time within the compute-heavy hydro solver. However, looking closer, the majority of time in the hydro solver is spent on local communication routines for filling guardcells and ensuring flux conservation. Similarly, the solvers for nuclear burning, flame, and turbulence spend the vast majority of time also filling guardcells. This indicates that our application spends most of its time in local MPI communication.



\begin{figure}[ht!]
    \centering
    \includegraphics[width=0.5\textwidth]{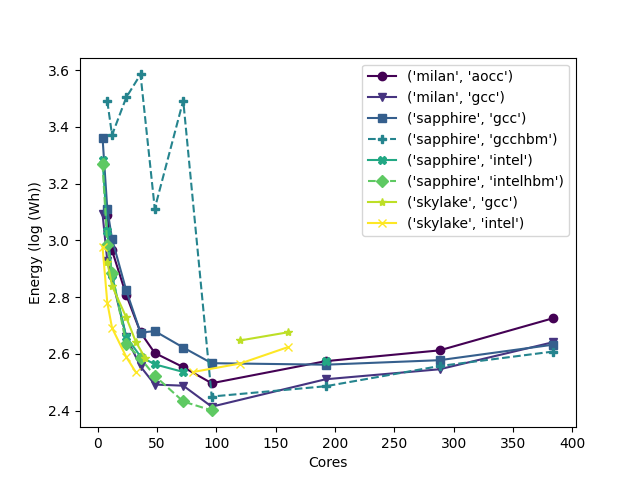}
    \vspace{-15pt}
    \caption{Energy consumption for strong-scaling study on all configurations available on SeaWulf from 4 to 384 cores. Missing data points are explained above.}
    \label{fig:small_energy}
\end{figure}

As expected in this strong-scaling study, we reach a point where the decrease in runtime is no longer associated with a decrease in energy consumption. As shown in Figure~\ref{fig:small_energy}, 
this minimum in energy consumption is reached for all configurations at 96 cores or 192 cores. In utilizing multiple nodes, the additional energy cost begins to outweigh the runtime improvement.

Overall, this strong-scaling study outlines the ability for FLASH to scale on a diverse set of configurations and outlines a clear region of minimal energy consumption from the 1-2 node mark. Upon reaching 3-4 nodes, it is clear that work-starvation is beginning to take effect, showcasing a need for optimization in our supernova problem and the FLASH code.


\subsection{Ookami}
\label{sec:ookami}
The relatively small memory per node of A64FX meant that it took some adjusting to fit our problem into memory. The configuration that worked was to set \texttt{maxblocks} to 50 and run on 36 out of 48 cores per node. This achieved a balance of keeping within the available memory per node while keeping \texttt{maxblocks} large enough to have room for grid refinement. Our application required a minimum of 21 {\color{black}A64FX-700} nodes (756 cores), and we scaled this across most of Ookami to 128 nodes (4608 cores).

\begin{figure}[ht!]
\centering
    \includegraphics[width=0.5\textwidth]{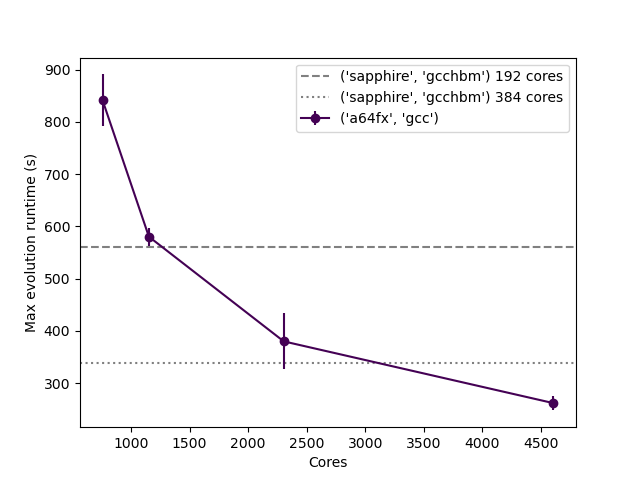}
    \includegraphics[width=0.45\textwidth]{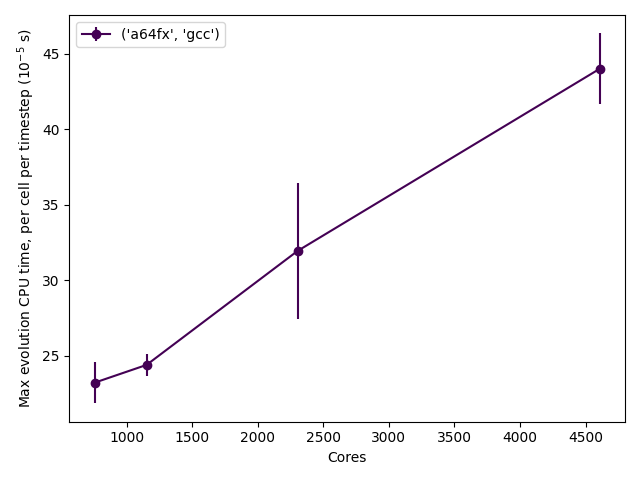}
  \vspace{-15pt} \caption{\label{fig:scaling_ookamigcc} Top panel: Strong-scaling study on {\color{black}A64FX-700}. Reported is the max evolution runtime. {\color{black} The two horizontal gray lines show the runtime for SPR+GCC with HBM for 192 (top, dashed line) and 384 (bottom, dotted line) cores,
the points for which would be off this scale.}
Bottom panel: The maximum evolution per-cell per-timestep CPU time.}
\end{figure}

Figure~\ref{fig:scaling_ookamigcc} shows the strong scaling of our application on {\color{black}A64FX-700}. {\color{black}A64FX-700} still shows good strong scaling after two doublings. {\color{black} The performance results for SPR with the GCC rather than Intel compiler are compared, in order to provide a more on par comparison to A64FX-700 with the GCC compiler. A64FX-700 performs almost as well as the 192 core (2 node) SPR+GCC with HBM runtime when using 1152 cores (32 nodes).} However, {\color{black}A64FX-700} does 10 times worse than any other architecture in the MCCTT metric. This is because 10 times as many cores are needed on A64FX to achieve the same performance as on other architectures.

One potential reason for the increasingly worse MCCTT is as the number of cores increases, more and more time is spent completing the global communication patterns of the gravity solver, rather than the compute-heavy and local communication patterns of the hydro solver. This indicates that tapering of scaling on Ookami is indeed driven by the increasing communication overhead.

\begin{figure}[ht!]
    \centering
    \includegraphics[width=0.5\textwidth]{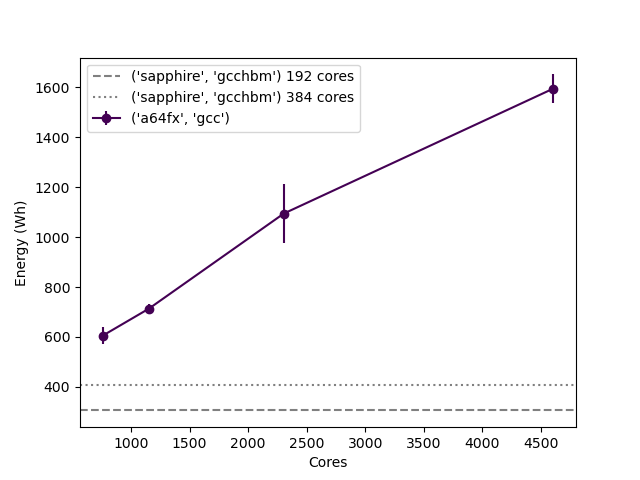}
    \vspace{-15pt}
    \caption{Energy consumption on {\color{black} A64FX-700 The two horizontal gray lines show the energy consumption for SPR+GCC with HBM for 192 (bottom, dashed line) and 384 (top, dotted line) cores,
the points for which would be off this scale.} }
    \label{fig:energy_ookamigcc}
\end{figure}

{\color{black} Figure~\ref{fig:energy_ookamigcc} shows the energy consumption on Ookami. The power efficiency of A64FX-700 -- only $\sim$ 120 W per node vs the $\sim$ 950 W per node of a full-subscription SPR node -- is overwhelmed by the sheer number of nodes needed to fit our problem in memory. Looking at points with comparable runtimes (192 cores of SPR and 1152 cores of A64FX-700, 384 cores of SPR and 2304 cores of A64FX-700), Ookami's A64FX-700 use 1.5$\times$ more energy than SeaWulf's SPR for our test application.} 

\section{HBM vs DDR5} \label{sec:hbm}
\begin{figure}[ht!]
\centering
    \includegraphics[width=0.49\textwidth]{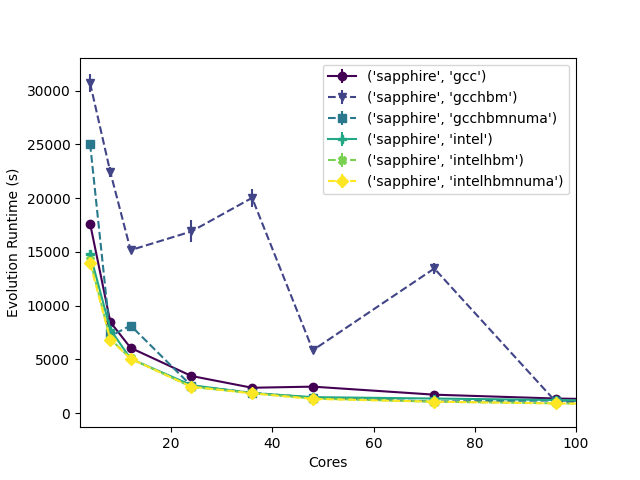}
    \includegraphics[width=0.49\textwidth]{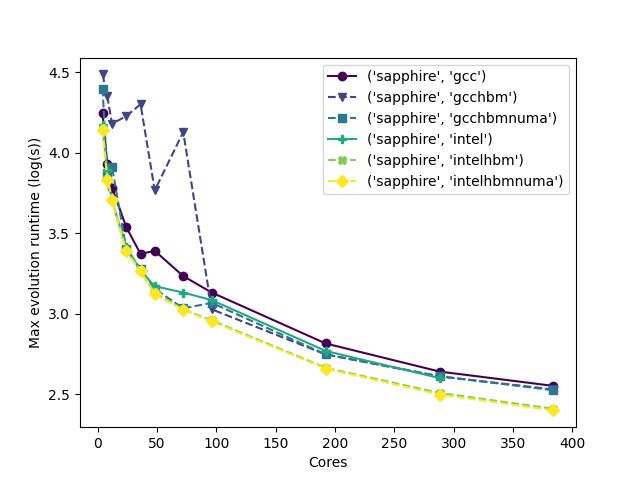}
    \vspace{-15pt} \caption{\label{fig:scalingHBM} Top panel: Strong-scaling studies on SPR from 4 to 96 cores (single node). Missing data points are explained above.
Bottom panel: The same strong-scaling study but expanded out to 384 cores and expressed as a log plot.}
\end{figure}

To better understand any benefits provided by HBM, we investigated our test problem on two architectures with this feature: SPR and A64FX. As discussed in Section~\ref{sec:1node}, using HBM with the SPR nodes produced some interesting behavior with the GCC compiler.
As shown in Figure~\ref{fig:scalingHBM}, enabling HBM does not provide a significant speedup for our application, which agrees with our assessment that FLASH is compute-heavy. When scaling out to multiple nodes, which allows the whole problem to fit within the HBM, the runtime difference between using only DDR5 and using HBM actually narrows.

\begin{figure}[ht!]
\centering
    \includegraphics[width=0.5\textwidth]{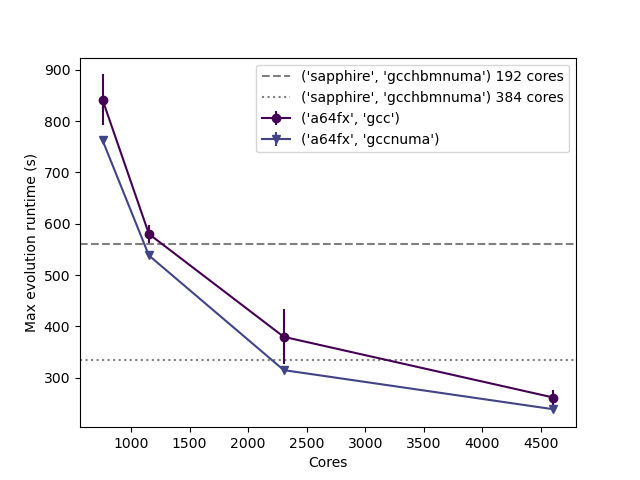}
    \includegraphics[width=0.5\textwidth]{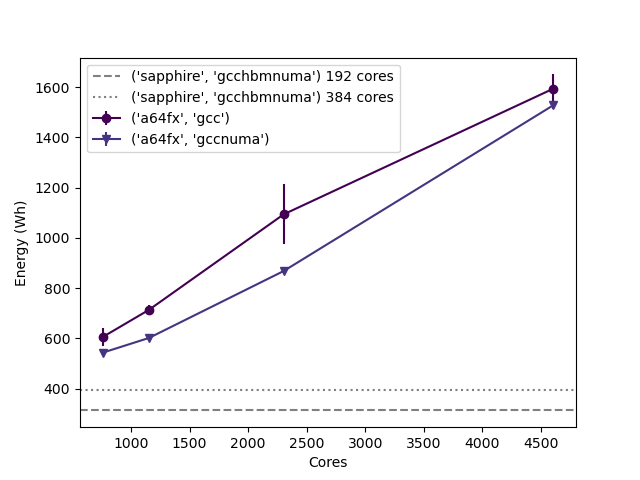}
  \vspace{-15pt} \caption{\label{fig:ookamiHBM} {\color{black} Max evolution runtime (top panel) and energy consumption (bottom panel) on A64FX-700 with and without \texttt{--map-by numa}. The two dotted grey lines show the runtime for SPR+GCC with HBM and the same directive for 192 (dashed line) and 384 (dotted line) cores, the points for which would be off this scale.} }
\end{figure}

Enabling the \texttt{--map-by numa} MPI directive appears to have the strongest effect on the effectiveness of HBM, when using fewer than all available cores per node. On SPR with GCC+OpenMPI, we see strange behavior without the directive when running on less than 96 cores (per node), and not much change in runtime when using full subscription. On A64FX, where HBM is the only memory type, the standard deviation of the runtime{\color{black}, and therefore the energy consumption,} is much higher without the directive, as shown in Figure~\ref{fig:ookamiHBM}. As we only run with 36 of the 48 cores on A64FX, we also see a significant decrease in runtime{\color{black}, and therefore energy consumption,} when using the directive. This indicates that there is a substantial difference between the default mapping on these systems and the NUMA mapping. At first, we thought this might have to do with memory locale, as MPI collectives appear to take longer without \texttt{--map-by numa}. However, the output of \texttt{numactl -s} is identical in both cases, indicating that the MPI rank mappings and the preferred memory are the same, which left us baffled. The effects of \texttt{--map-by numa} with OpenMPI are still being investigated.

\section{Core-spread Analysis} \label{sec:corespread}
Our finding that decreasing the number of cores running on a node decreased the MCCTT
suggested conducting a study in which we kept the same number of total cores constant but began to spread them across multiple nodes. We dubbed this study the "Core-spread Analysis" and first conducted this study on the SPR-GCC configuration because, at the time, this was our fastest configuration that worked reliably. This study was conducted with 192 cores because that is where SPR-GCC most efficiently runs our 220GB problem. We then extended this study to SPR-GCC with HBM as well as adding a NUMA-mapping to the SPR-GCC run which we have found to be the most efficient MPI mapping.

\begin{figure}[h!]
\centering
    \includegraphics[width=0.5\textwidth]{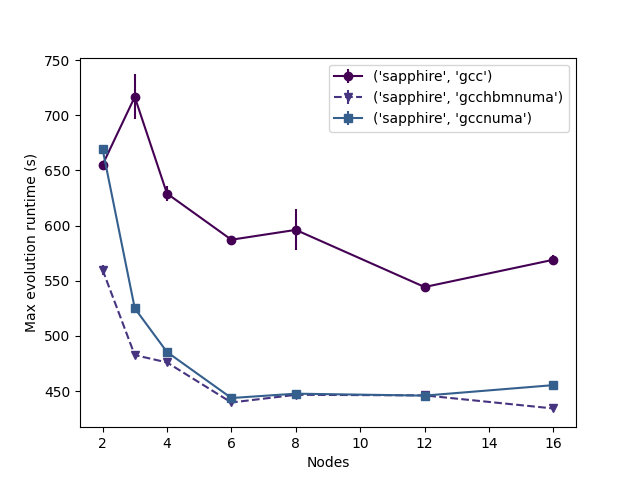}
    \includegraphics[width=0.5\textwidth]{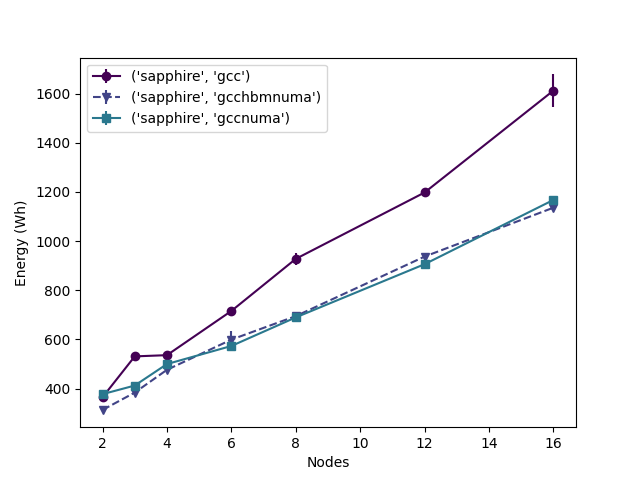}
  \vspace{-15pt} \caption{\label{fig:small_core_spread} Top panel: Core-spread analysis detailing a constant 192 cores spread across a variable number of nodes.
Bottom panel: Energy consumption for core-spread analysis at different node counts.}
\end{figure}

Figure~\ref{fig:small_core_spread} presents our core-spread analysis leading to a few key insights. First, it is important to note that choosing to map by NUMA region vs. using the default mapping causes a horizontal shift in where the minimum occurs. With default mapping, the minimum runtime occurs at 12 nodes. However, this minimum is shifted to 6 nodes when we map by NUMA on SPR+GCC with and without HBM. This shift in minima is important as the more nodes we use to spread our cores, the higher the energy consumption. It follows that mapping by NUMA region not only allows us to achieve lower runtimes, it also increases energy efficiency and creates a smaller node footprint in the optimal run configuration.

\begin{table*}[ht!]
\caption{Core-spread analysis for different architectures. Shown are the number of nodes and cores, the evolution runtime, the measured average core frequency, the measured average power for one node, and the total energy, per run. {\color{black} The average power for Ookami was collected with XDMoD.} }
\label{tab:CPUFreq}
\begin{tabular}{lcccrrrr}
\hline \hline
Architecture                     & \multicolumn{1}{l}{Total cores} & \multicolumn{1}{l}{Nodes} & \multicolumn{1}{l}{Cores/node} & \multicolumn{1}{l}{Runtime (s)} & \multicolumn{1}{l}{Frequency (Hz)}  & \multicolumn{1}{l}{Power (W)} & Energy (Wh)                        \\
\hline
\textbf{SPR HBM} & 192                             & 2                         & 96                                 & 563.6                           & 2.601 $\pm$ 0.005 & 916.9                         & 314.1 $\pm$ 3.7                     \\
                                 & 192                             & 3                         & 64                                 & 486.8                & 3.079 $\pm$ 0.242 & 876.9                         & 383.2 $\pm$ 5.6                      \\
                                 & 192                             & 6                         & 32                                 & 442.2                           & 3.498 $\pm$ 0.007 & 699.4                         & 570.6 $\pm$ 5.4                      \\
                                 \hline
\textbf{Milan}   & 192                             & 2                         & 96                                 & 866.0                           & 2.839 $\pm$ 0.063                     & 642.2                         & 354.3 $\pm$ 2.7 \\
                                 & 192                             & 3                         & 64                                 & 614.3                           & 2.877 $\pm$ 0.313                     & 577.4                         & 308.8 $\pm$ 1.6  \\
                                 & 192                             & 6                         & 32                                 & 600.2                           & 2.850 $\pm$ 0.433                     & 405.9                         & 442.4 $\pm$ 7.6  \\
                                 \hline
\textbf{A64FX-700}  & 756                             & 21                        & 36                                 & 762.5                          & 1.8                                 & 125.4 $\pm$ 4.4      &      609.9 $\pm$ 2.3              \\
                                 & 756                             & 27                        & 28                                 & 737.9                           & 1.8        &    119.5 $\pm$ 6.6      &     723.8 $\pm$ 2.0                              \\
                                 & 756                             & 63                        & 12                                 & 682.6                           & 1.8   &  130.0 $\pm$ 9.8   &    1705.1 $\pm$ 7.0      \\                        
\hline
\end{tabular}
\end{table*}

During the core-spread analysis study, we also experimented with other potential reasoning for why spreading our cores across multiple nodes can create a decrease in runtime. 
We selected SPR, Milan, and {\color{black} A64FX-700} for our tests. For each architecture we used the GCC compiler, OpenMPI, and the \texttt{--map-by numa} MPI directive, which as discussed in Section\ref{sec:hbm} affects runtime when using not fully-subscribed nodes. We kept \texttt{maxblocks} consistent because the same number of total cores are used for each architecture. To measure the CPU frequency, we polled \texttt{cpupower frequency-info} every 10 seconds for each CPU on a single node, over the course of all 7 runs. The average evolution runtime, average CPU frequency for one node, average power for one run for one node, and average total energy for one run are reported below in Table~\ref{tab:CPUFreq}.

On SPR, idle cores run at 3.5 GHz, while on Milan, idle cores decrease in frequency. For a fully-subscribed node, all cores more or less run at the same frequency, but as the number of cores per node decreases, there is more of a spread in CPU frequencies. This is shown by an increase in standard deviation. We discovered that when we spread the same number of cores across multiple nodes, we raise the clock speed on each CPU and this provides further speedup in runtime. The speedup is mostly due to this change in clock speed; however it is not enough to account for the full runtime difference. This can be explicitly seen on A64FX, which has a stable clock speed but still shows some runtime improvements when spreading across nodes. This smaller speedup can be attributed to the increase in bandwidth provided by adding another node, among other factors. {\color{black} It is interesting that despite the communication bottleneck, spreading across nodes and therefore increasing the amount of inter-node communication produces a speedup.} However, given a certain number of nodes, it is always faster to use all available cores. Therefore, while spreading our application across nodes is interesting and provides insight, it is not the most practical way of running our application.

\section{Conclusions and Future Work} \label{sec:conclusions}
Our application quickly becomes communication-bound even within a single node, and spends most of its time filling the guard cells of each block rather than performing computation. Enabling HBM does not give us much of a speedup, indicating that our application is not memory-bound.

SPR gives the fastest time to solution by expending the least energy. Milan is not far off, but exhibits worse scaling behavior with increasing numbers of cores. {\color{black} A64FX-700} can match the performance of other architectures, but requires the use of 10 times as many nodes, which overwhelms its power efficiency advantages. The native Intel compiler is faster for the Intel architectures, but interestingly this is not the case for the AOCC compiler on Milan.

The increase in MCCTT when scaling within a node, shown in Figure~\ref{fig:small_dean}, can be partially attributed to the decrease in processor speed as more and more cores are used on the node. This reaches an equilibrium when the node is full and all the cores are running at the same lower frequency. However, the decrease in processor speed alone isn't enough to account for the single-node increase in MCCTT shown in Figure~\ref{fig:small_dean}. As our application is communication-bound, the addition of more cores increases the communication overhead and therefore the MCCTT. Interestingly, enabling HBM seems to ameliorate the slope of increase.

For a given number of processors, spreading them across nodes can produce a faster runtime than using the minimum required number of nodes. This is interesting but may not be practical, as the fastest time to solution on a given number of nodes comes from filling the node.

Profiling reports on our supernova problem reveal a significant need for better vectorization and parallelization of the underlying code. Making these optimizations will allow for better overall performance. Currently, FLASH is not well vectorized because of both the way block data is structured and due to the fact that iterative solvers, such as the hydro solver, will not take the same amount of time or perform the same number of iterations to converge. Vectorization of FLASH is difficult, and this massive undertaking is currently underway by the FLASH-X team in their goal of performance portability. This effort will also be helpful in improving FLASH's threading capcbilities. Currently, only a few solvers in FLASH are threaded (< 10\% of our application), and initial investigation indicates that enabling threading gives much less of a performance improvement than simply using the cores as additional MPI ranks.


However, as we are communication-bound these improvements will not address the main bottleneck. One avenue we could explore is communication buffer packing to minimize latency and improve communication efficiency, but a study with a similar AMR scheme did not see a benefit to this on CPUs~\cite{parthenon}. Certainly a more detailed study of FLASH's communication patterns is required to improve the performance.

{\color{black} Fugaku may offer a solution, with an improved hardware and interconnect than that of Ookami. Fugaku's A64FX-1000 has a 2.0 GHz clock speed with a 2.2 GHz boost mode, 20\% faster than that of Ookami. There are also 4 extra ``assistant'' cores per node to help with I/O. The custom Tofu-D interconnect may also be beneficial for our application's communication bottleneck. As the 32 GB of memory is the same, we will still need to use a large number of nodes, but the improvements in the hardware and interconnect may decrease the runtime, and subsequently the energy consumption. If, through a combination of features, Fugaku can decrease the energy consumption by 30\%, then it will be on par with SPR, while perhaps providing even better performance.}

\begin{acks}
The authors are grateful to Dean Townsley for insightful discussion, especially with regards to data analysis and CPU frequencies. We also thank Nikolay Simakov {\color{black} and Joseph White} for help obtaining data from Ookami's XDMoD.
This work was supported in part by the US Department of Energy (DOE) under grant DE-FG02-87ER40317 and by the National Science Foundation (NSF) under grant 1927880. The SeaWulf computing system was made possible by NSF grants (\#1531492 and \#2215987) and the Ookami computing system by a \$5M NSF grant (\#1927880). FLASH was developed in part by the DOE NNSA ASC- and DOE Office of Science ASCR-supported Flash Center for Computational Science at the University of Chicago. This research has used NASA’s Astrophysics Data System Bibliographic Services.

\end{acks}

\bibliographystyle{ACM-Reference-Format}


\appendix

\end{document}